\documentclass[10pt,conference]{IEEEtran}

\newcommand{\forcameraready}[1]{}

\usepackage[
  pass,
]{geometry}

\usepackage{balance}
\usepackage[T1]{fontenc}
\usepackage[utf8]{inputenc}
\usepackage{lmodern}
\usepackage{color}
\usepackage{paralist}
\usepackage{lipsum}
\usepackage{booktabs}
\usepackage{multirow}
\usepackage{tabularx}
\usepackage{fixltx2e}
\usepackage{microtype}
\usepackage{tikz}
\usepackage[normalem]{ulem}
\newcommand{\subparagraph}{} 
\usepackage{mathptmx}
\usepackage[cmex10]{amsmath}
\usepackage{amssymb}

\usepackage[labelfont=bf]{subcaption}
\usepackage{listings}
\usepackage{xspace}

\definecolor{mygray}{rgb}{0.5,0.5,0.5}

\usepackage[hidelinks]{hyperref}
\usepackage{xspace}

\def\natural{tailored\xspace}

\newcommand{\ourtitle}{Tailored Mutants Fit Bugs Better}

\hypersetup{
  pdftitle={\ourtitle},
  pdfauthor={\forcameraready{Miltiadis Allamanis, Earl T. Barr, Ren\'e Just, and Charles Sutton}},
  plainpages=false,
  pdfpagelabels
}

\newcommand{\defref}[1]{\hyperref[#1]{Definition~\ref*{#1}}}
\newcommand{\algref}[1]{\hyperref[#1]{Algorithm~\ref*{#1}}}
\newcommand{\lineref}[1]{\hyperref[#1]{Line~\ref*{#1}}}

\newcommand{\cameraready}[1]{}

\newcommand{\etal}{\hbox{\emph{et al.}}\xspace}
\newcommand{\eg}{\hbox{\emph{e.g.}}\xspace}
\newcommand{\ie}{\hbox{\emph{i.e.}}\xspace}

\newcommand{\etc}{\hbox{\emph{etc.}}\xspace}
\newcommand{\vs}{\hbox{\emph{vs.}}\xspace}

\newcommand{\ngram}{$n$-gram\xspace}

\newcommand{\naturalize}{\textsc{Naturalize}\xspace}

\usepackage{algorithmicx}
\usepackage{algpseudocode}

\newlength{\emstr}
\newcommand{\boldpara}[1]{%
  \smallskip%
  \par\noindent\textbf{\textit{#1}}\hspace{\emstr}
}%

\newenvironment{squishlist}
{
 \begin{list}{$\bullet$}
  { \setlength{\itemsep}{0pt}
     \setlength{\parsep}{3pt}
     \setlength{\topsep}{3pt}
     \setlength{\partopsep}{0pt}
     \setlength{\leftmargin}{1.5em}
     \setlength{\labelwidth}{1em}
     \setlength{\labelsep}{0.5em} } }
{  \end{list}  }

\begin{document}

\title{\ourtitle}

\author{\IEEEauthorblockN{Miltiadis Allamanis}
\IEEEauthorblockA{School of Informatics\\
University of Edinburgh\\
Edinburgh, EH8 9AB, UK\\
m.allamanis@ed.ac.uk}
\and
\IEEEauthorblockN{Earl T. Barr}
\IEEEauthorblockA{Dept. of Computer Science\\
University College London\\
London, UK\\
e.barr@ucl.ac.uk}
\and
\IEEEauthorblockN{Ren\'e Just}
\IEEEauthorblockA{Dept. of Computer Science\\
University of Massachusetts\\
Amherst, MA, USA\\
rjust@cs.umass.edu}
\and
\IEEEauthorblockN{Charles Sutton}
\IEEEauthorblockA{School of Informatics\\
University of Edinburgh\\
Edinburgh, EH8 9AB, UK\\
csutton@ed.ac.uk}
}

\maketitle

\begin{abstract}

%
%
%
Mutation analysis measures test suite adequacy, the degree to which a test suite
detects seeded faults (mutants): one test suite is better than another if it
detects more mutants. The effectiveness of mutation analysis rests on the
assumption that mutants are coupled with real faults---that is, mutant detection
is strongly correlated with real fault detection.  The work that validated
this assumption also showed that a large portion of defects remain out of reach.

We introduce tailored mutation operators to reach and capture these defects.
Tailored mutation operators are built from and apply to an existing code base
and its version history.  They can, for instance, identify and replay errors
specific to the project for which they are tailored.  As our point of
departure, we define tailored mutation operators for identifiers, which
mutation analysis has largely ignored, because there are too many ways to
mutate identifiers.  Evaluated on the Defects4J data set, our new mutation
operators allow mutation analysis to create mutants coupled to 14\% more
faults, compared to traditional mutation operators.

These new mutation operators, however, quadruple the number of mutants,
exacerbating the problem of mutant selection.  To combat this problem, this
paper proposes a new approach to mutant selection that focuses on the location
at which to apply mutation operators and the unnaturalness of the mutated code.
The results demonstrate that the location
selection heuristics produce mutants more closely coupled to real faults for a
given budget of mutation operator applications.

In summary, this paper defines and explores tailored mutation operators,
advancing the state of the art in mutation testing in two ways:  1) it
suggests mutation operators that mutate identifiers and literals, extending
mutation analysis to a new class of faults and 2) it demonstrates that
selecting the location where a mutation operator is applied decreases the
number of generated mutants without affecting the coupling of mutants and real
faults.


\end{abstract}

\begin{IEEEkeywords}
mutation testing; submodular optimization; code naturalness

\end{IEEEkeywords}

\section{Introduction}
\label{sec:intro}

Mutation analysis is a well-established approach to measure the effectiveness of a test
suite. The \emph{mutation score}, which measures a test suite's ability to distinguish a
program under test from many small syntactic variations (\emph{mutants}), is a proxy
metric for that test suite's ability to detect real faults. These mutants are created by
applying well-defined \emph{mutation operators}, which are program transformations that
systematically inject small artificial faults into the program under test. Examples of
such mutation operators include replacing an arithmetic operator, manipulating a branch
condition, or deleting a statement.

The usefulness of this proxy metric rests on the assumption that a test suite's ability to
detect mutants is strongly correlated with its ability to detect real faults, and prior
work provided empirical evidence for the existence of such a
correlation~\cite{andrews2005mutation,just2014mutation}.
Indeed, one fundamental assumption of mutation analysis is the existence of the \emph{coupling
effect}~\cite{lipton1978mutation}. A real fault, which is usually complex, is coupled to a set of
mutants, which are usually simple, if a test that detects all the mutants also detects the real
fault.  Just \etal empirically showed that such a coupling effect indeed exists between
\emph{traditional mutants}, \ie, mutants generated by commonly used mutation operators,
and most real faults~\cite{just2014mutation}. However, they also identified inherent
limitations of mutation analysis: 27\% of real faults were not coupled to any mutant due
to the lack of suitable mutation operators.
This suggests a need to introduce new mutation operators that are designed
to be coupled to these uncovered faults.

However, introducing new mutation operators can be problematic, because
mutation analysis suffers from scalability problems due to the large number of mutants
that can be generated---even for mid-sized programs, and selective mutation is the most
common approach to control the computational costs. A number of empirical studies conclude
that randomly selecting mutants from a pool of generated mutants is at least as effective
as selecting a particular set of mutation
operators~\cite{zhang2010selective,gopinath2016limits}. Moreover, Kurtz \etal
showed that no single selective mutation approach works reasonably well for all programs,
and they argue that a good mutant selection strategy depends on the program under
test~\cite{kurtz2016selective}.

This paper aims to improve the effectiveness of mutation analysis by introducing
new mutation operators that are tailored to the specific program under analysis.
By \emph{tailored} we mean that the set of code transformations produced
by the mutation operator is specific to the project in question,
in the sense that the mutants are built from and apply to an existing code base
and its version history.  For example, a tailored mutation operator might use literals
that occur elsewhere in the project, insert calls to project-specific
APIs,  insert  code fragments from
elsewhere in the project, or even identify and replay errors
from the project's version history.
To the best of our knowledge, we are the first to introduce such
tailored operators.
More specifically, in this work, we introduce tailored mutation operators for
identifiers and literals: Our operator for identifiers replaces
one identifier with another that is of the correct type and is in scope,
and our operator for literals replaces a literal with one that is used elsewhere in the project.

These tailored operators are extraordinarily prolific.
A single operator can in some cases generate thousands of mutants
at a single lines of code, and tailored operators
generate on average \emph{4 times} more mutants than a state of the art
set of traditional mutation operators.
Because each individual mutation operator is so prolific, selecting mutations
at the operator level is not effective.
Instead we introduce two new methods for selecting individual mutants.
First, we introduce a novel approach to selecting locations at which to mutate
based on \emph{submodular optimization} that intuitively attempts to select locations
that are as far apart as possible in the program's control flow graph.
Submodularity is a well-known property of set functions, analogous to convexity
in continuous functions, that has seen wide applications in economics,
operations research, and artificial intelligence, but seems to have received
less attention in software engineering.
Second, we introduce a method for ranking individual mutants based on naturalness.
Following the insight of Ray \etal \cite{ray2015naturalness}
that bugs are often unnatural, we rank highest unnatural mutants, that is,
those that have low probability according to an $n$-gram language model.

 Our empirical evaluation studies the
improvements in effectiveness and efficiency these operators bring to mutation
testing, using 253 real faults and 659,326 mutants. The results show that the
new mutation operators are coupled with 67\% of the defects, increasing the overall
ratio of coupled defects to 80\% when combined with traditional mutants. In addition, our novel mutant selection approach, for a
given target effectiveness, can greatly reduce the execution budget required for mutation testing. For example,
to find mutants that are coupled to 70\% of the defects, naive random selection of mutants requires
on average twice as many mutants as our selection method to reach this level of effectiveness.

The practical implications for mutation testing include
\begin{squishlist}

  \item When focusing on a specific scope of a method or a line to detect bugs,
   using a combination of traditional and tailored mutation operators will
   improve the effectiveness of the mutation analysis;

  \item When using either traditional mutants and tailored mutants, methods
  that select mutants based on their location can provide better efficiency
  for a given budget.

\end{squishlist}

\section{Approach}
\label{sec:approach}

Traditional mutation operators have proven their utility and enjoyed industrial
adoption.  One reason for their success is their universality: to increase
their applicability to programs across many application domains and even across
programming languages, these operators have tended to focus on local,
fine-grained changes to symbols with near universal meaning, like relational
operators.  Despite their success, previous work~\cite{just2014mutation,gopinath2014similarity}
has shown that traditional mutants are dissimilar to real faults and, more importantly,
are not coupled to important classes of defects.

Motivated by Kurtz \etal~\cite{kurtz2016selective} who found that an effective set of
mutation operators depends on the program, and Just \etal~\cite{just2014mutation}
who found that new stronger mutation operators are needed, such as faults
that have a ``similar (library) method called'', ``specific literal replacement'' \etc,
we extend traditional mutation operators with tailored operators.
These operators are tailored to a particular program, yet retain universality
in the sense that they are automatically generated for an input program.  One
class of these operators can be extracted from a project's version history.
For instance, one might identify bug-introducing commits and extract a mutation
operator from them.  A canonical class of tailored operators is literal
mutators.  Here a na\"ive choice of replacements is unbounded.

To square the circle of realizing finite customization, we turn to the notion
that software is natural, that it is locally regular and predictable~\cite{hindle2012naturalness}.
Using this observation, Ray \etal~\cite{ray2015naturalness} found that less
natural code tends to be more ``buggy''.
Naturalness allows us to rank alternatives when defining tailored mutation operators.
When defining a literal replacement operator, for instance, we rank our choices
in terms of the ones that are the most unnatural, or most surprising, to a
language model and drop the rest.  It also allows us to constrain where we apply
our mutation operators and rank the operators to apply at those locations.  In
each of these cases, the intuition is that highly-ranked mutants are more likely to be
similar to
mistakes a developer might make and therefore produce mutants that are coupled
with some of the defects that traditional operators do not reach.

\subsection{Terminology}

%

As is standard, a mutation operator is a rewriting rule.  A mutant is a program
variant produced by the application of a mutation operator.  We consider only
mutants produced by a single application of a mutation operator, not those
produced by $k > 1$ applications, so called higher order
mutants~\cite{jia2009hom}.  As a result, we sometimes use mutant and the
application of a mutation operator interchangeably in the following.  A
\emph{tailored mutant} is a mutant produced by a tailored mutation operator.

We use the term ``program location'' or simply ``location'' to denote a node in a
fine-grained control flow graph (CFG), where, instead of a basic block, each
node represents a single statement.  At the source code level, each program location refers to one or
more code tokens.  In the current implementation of our system,
we generate CFGs at an intraprocedural level, which means that we have
one CFG for each method in the program. A sample CFG can be seen in \autoref{fig:sampleCfg}.

\subsection{Tailored Mutation Operators}

Tailored mutation operators can be built from commits or from recurrent
syntactic patterns; they can perturb API protocols; they can operate on ASTs. A
concrete example of an AST operator would be one that replaced function
overloads, like \lstinline+return solve(min, max)+ with
\lstinline+return solve(f,min,max)+ where \lstinline+f+ is type compatible
and in scope.  As proof-of-concept, we consider tailored operators that replace
only one token at a time, like identifier names.  A tailored operator can
replace an identifier name with another name used in similar contexts or even
simply another name in scope.  In this paper, we introduce identifier and
literal replacement operators.  These operators do not preserve semantics, but
respect the type system and produce compilable mutants.

Our identifier operator applies to function calls and uses of variables,
parameters and fields.  When applied to a function call, it replaces the function name
with the name of another function that has the same signature, respecting the scope of the receiver
object of the original method call.  When applied to uses of variables,
parameters, or fields, it replaces the name at that usage with another name of a variable,
parameter, or field that is in scope and compatible.
A replacement variable name is compatible if it is
type-compatible while a replacement field name is compatible if the alternate
name is another field of the containing object.

Traditional mutation operators include an operator for literals that selects from
a small, fixed set of type-compatible literals, such as -1, 0, and 1 for an integer
literal, or the empty string or \lstinline+null+ for a string literal. While such a
pre-defined set is effective in some cases, it misses specific replacements that cause
subtle defects and, moreover, it generates a large number of trivial mutants. For example,
replacing a dereferenced string literal with a null reference or a number literal that represents an
array index with a negative number, raises an exception as soon as that mutant is
executed, rendering it as trivial.

Mutating literals defines a potentially
unbounded set of mutants; we exploit the naturalness of software to truncate and
rank the infinite possibilities, by considering literal replacements that
are common within a code base.  To mutate literals, we use a project-specific,
context-specific approach by mutating literals and variable usages to literals
that are commonly used within that code base at a similar context.  This process
begins by looking for all the trigrams in the rest of the code base that have the
same prefix as the current token being mutated. This step ensures that
common literals of the code base are added to the list of candidates.
Additionally, for literal mutation we add as candidates all primitive variables
of the correct type. Finally, we remove redundant target numeric literals that
have equivalent values (\eg \lstinline+-0+ and \lstinline+0+ or
\lstinline+10e-1+ and \lstinline+0.1+).

Our evaluation shows that these tailored mutation operators capture a wide variety of
real defects that humans introduce into the code.  They also produce vastly more
mutants, exacerbating the cost of mutation analysis.  Next, we discuss how we
address this problem by judiciously deciding where to apply which operators.

\subsection{Selecting Locations}

To exploit our new mutants, we must sample them efficiently.  Here we explore
selection policies that allow
us to select a smaller set of mutants, without sacrificing their effectiveness.
Since the number of mutants within a program location are usually limited, we first
explore two methods for selecting mutation program locations. Selecting these locations
allows us to pick a smaller set of mutants. As a baseline, we use the effectiveness
of a \emph{fully random} selection process. This process randomly selects a subset of the mutants
from the pool of all available mutants.

\boldpara{Random Location First}
The first policy that we consider is to select a random program location first and then
to select a mutation operator within that location. This two-stage random
process removes any bias because of an imbalance
on the number of mutants across locations, \ie it has the effect that it
down-weights mutants that appear in more ``crowded'' locations. To
compute the effectiveness of this selection policy, we resort to a Monte
Carlo simulation and average the results across simulations.

\boldpara{Minimum CFG Distance Selection}
Selecting locations at random is reasonable,
but may not be optimal.
We are interested in mutating as many relevant code paths as possible to make
sure that all possible cases have been mutated. Additionally,
we need to pick locations that maximize the marginal
utility of each mutation. If we chose two nearby mutants there is a
higher chance that the ``fates'' of those two mutants are correlated.
Both of these considerations suggest that we should pick highly diverse locations
to maximize the expected utility for each mutant execution.
To achieve this we
explore an optimization-based approach for selecting mutants.

One reasonable heuristic is to choose mutants in locations
so as to minimize the total distance between each CFG node (\ie location)
to the closest node with a selected mutant.  In other words, we wish to place mutants
such that they are near as many statements as possible
in the CFG. This heuristic
implicitly selects mutants to increase coverage metrics.
We formulate this as an optimization problem over sets, \ie,
let $\mathbb{L}$ be the set of all possible mutant locations at which at least
one mutation is possible.  Then we define an objective function
$O : 2^{\mathbb{L}} \rightarrow \mathbb{R}$
over a potential set of locations
\begin{equation}
	 O(L) = \sum_{g \in G}\sum_{n \in N_{g}} \min_{l \in L} d(n,l),
\end{equation}
and select a set $L^{*}_{\kappa} \subseteq \mathbb{L}$
of CFG nodes (locations) by solving the optimization problem
\begin{equation}
\label{eq:distanceOptimization}
	L^{*}_{\kappa} = \arg\min_{L} O(L)
	\end{equation}
subject to the constraint $|L|=\kappa$, where $\kappa$ is a user-selected
maximum number of mutants,
$G$ is the set of CFGs for all methods in the program, and $N_g$ is the set of all locations (nodes) in the CFG $g$.
This optimization essentially selects $\kappa$ CFG nodes (in $L^{*}_{\kappa}$) to minimize
the average distance between all nodes to their closest node in the set of selected nodes.

This optimization problem especially convenient because it is
\emph{submodular}~\cite{bach:submodular}\footnote{Many more references and tutorials
are available at \url{http://submodularity.org} by Andreas Krause
and Carlos Guestrin.}.
Submodularity is a diminishing returns property, which says intuitively that
if we consider adding one more location $l$ to an existing set $L$,
the benefit of adding $l$ diminishes as the set $L$ gets bigger.
Submodularity has found application in wide range of similar problems
to ours,
including placement of sensors in sensor networks~\cite{krause08near} and
identifying the most influential nodes in a social network \cite{kempe2003maximizing}.
More formally, let $T$ be an arbitrary set and let $f : 2^T \rightarrow \mathbb{R}$
map sets $S \subseteq T$ to the real numbers. We say that $f$ is \emph{submodular} if
for all $S \subseteq S' \subseteq T$ (implying
$\left|S\right| \leq \left|S'\right| \leq \left|T\right|$) and for all $x \in T - S'$, we have
\begin{equation}
	f(S \cup \{ x \}) - f(S) \geq f(S' \cup \{ x \}) - f(S').
\end{equation}
which implies that adding an extra element $x$ to a larger set $S'$ yields a smaller
increase compared to adding it to a smaller set $S$.
More precisely, we show that the function $-O$ is submodular.
For any $L \subseteq L' \subseteq \mathbb{L}$ and $l' \in \mathbb{L}$ with $l' \not\in L'$, first consider a
 single CFG $g \in G$ and a single node $n \in N_g$.
Then we have
\begin{align*}
 \left[\min_{l \in L} d(n, l) - \min_{l \in L \cup \{l'\}} d(n,l) \right]
	&=  \min\left\{0, \min_{l \in L} d(n,l) - d(n,l') \right\} \\
	&\geq \min\left\{0, \min_{l \in L'} d(n,l) - d(n,l') \right\} \\
	 &= \min_{l \in L'} \left[ d(n,l) - \min_{l \in L' \cup \{l'\}} d(n,l)\right].
\end{align*}
Now we can simply sum this inequality over all CFGs $g \in G$
and all nodes $n \in N_g$ to obtain the
desired result, namely, that
$O(L) - O(L \cup \{ l \})  \geq O(L') - O(L' \cup \{ l \}).$

Solving the optimization problem \eqref{eq:distanceOptimization}
requires maximizing a submodular function under a cardinality constraint. Although this problem is NP-hard,
we resort to a greedy approximation, as is typical in the literature,
for which strong approximation guarantees exist (see \cite{krause:submodular} for details).
At each step $i$ we greedily pick the location that \emph{minimizes} the sum of the
distances of each CFG node to the nearest node that belongs to the set of selected nodes.

\begin{figure}[t]
	\centering
	\includegraphics[width=0.4\columnwidth]{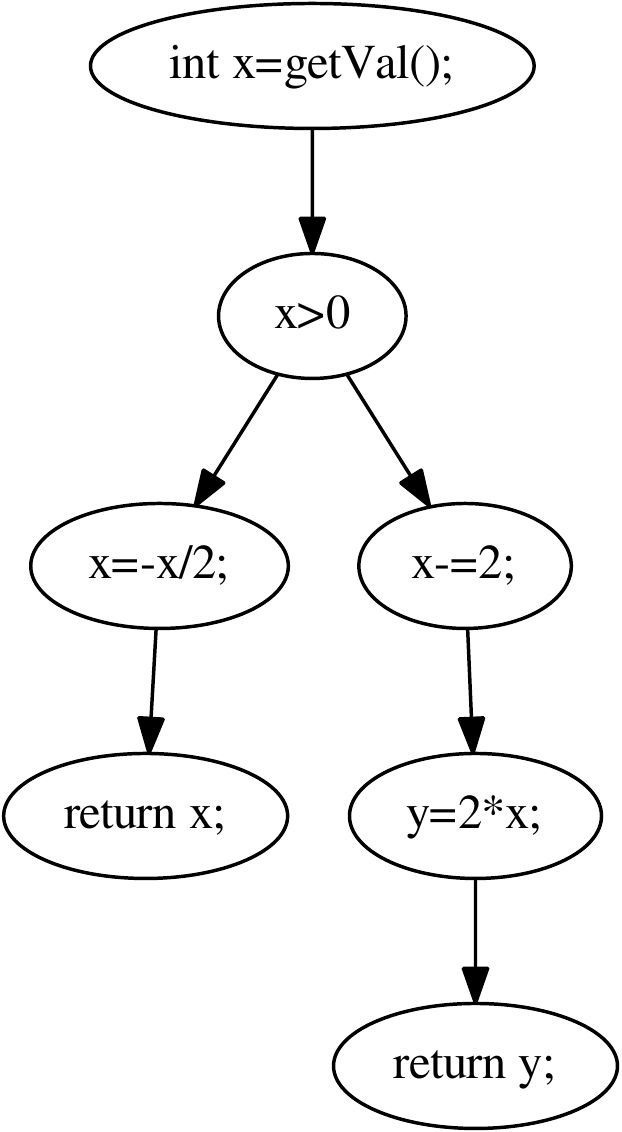}
	\caption{An example control flow graph similar to those used in this work.
  Each node represents a single statement. The optimization objective
  \eqref{eq:distanceOptimization} tries to pick $M$ nodes (locations) to minimize
  the sum of the distances of all nodes to the nearest selected nodes.}
	\label{fig:sampleCfg}
\end{figure}

To measure the distance $d(s,t)$ between two CFG nodes $s$ and $t$,
we use the smaller of the length of shortest path from $s$ to $t$,
or the length of shortest path from $t$ to $s$ (recall that the
CFG is directed).
Note that the distance between two nodes
in mutually exclusive branches is infinite, since there is
no path connecting them. Therefore,
optimizing our objective implicitly maximizes branch coverage. Additionally,
this selection process implicitly follows the intuition that
branching nodes (\eg \lstinline+if+ conditionals) should be
selected early, since selecting such a node reduces the distance to nodes within both branches.

To illustrate \autoref{eq:distanceOptimization}, consider the
example graph in \autoref{fig:sampleCfg}. To find $L^*_1$ we consider all
nodes in the graph and select $L^*_1=\{$\lstinline+x>0+$\}$. The sum of the distances of all
nodes to the \lstinline+x>0+ node is $O(L^*_1)=10$, which minimizes $O$ subject to $\kappa=1$. If
we were to pick a node in any of the branches during the first step, $O$ would have been infinite
since there is no path connecting nodes of two separate branches. Next,
to find $L^*_2$, we select  $L^*_2=\{$\lstinline+y=2*x;+$\}\cup L^*_1$ achieving $O(L^*_2)=6$.  The
next step would include selecting either node \lstinline+x=-x/2;+ or node
\lstinline+return x;+ to reduce $O(L^*_3)$ to 4.
Note that although our optimization objective does \emph{not} explicitly
optimize for branch coverage or branching conditions, it implicitly covers our
intuition that conditions should be mutated early and that high branch coverage
should be achieved. Additionally, note that statements within basic
blocks that contain more statements will be preferred to blocks with fewer statements.
Finally notice that because of the submodularity of $-O$, adding extra nodes to
the set of selected nodes has diminishing returns ($O(L^*_1)-O(L^*_2)=4$, $O(L^*_2)-O(L^*_3)=2$, $O(L^*_3)-O(L^*_4)=1$, \etc).

\subsection{Picking Mutation Operators at Location}
Once a location is selected, we still need to pick the ``best'' mutant from the set of
all available mutants at the specified location. One obvious approach is to
pick a random mutant. We use this policy as our baseline.

\boldpara{Scoring Mutation Operators by Naturalness}
Recent work in \emph{naturalness of code}~\cite{hindle2012naturalness} allows us to measure how
probable a specific token is given information from the rest of the code within
each project. Furthermore, Ray \etal~\cite{ray2015naturalness} has recently
found that unnatural code is on average more buggy. Therefore, mutating code
to make it \emph{less} natural implies a clear way of generating bugs.

Inspired from
the \naturalize framework~\cite{allamanis2014learning,allamanis2015suggesting}
that used probabilistic models of source code to suggest alternative names for
variables, methods and classes, we use the same technique
to score tailored mutation operators. Note, that although Allamanis \etal
suggest names that \emph{increase} the naturalness of the code, here we prioritize
mutants the \emph{decrease} its naturalness.
 To measure the
``naturalness'' of each mutation, we use an \ngram language model.
 A language model (LM) is a probability distribution over
strings.  Given any string $x = x_0, x_1 \dots x_M$, where each $x_i$ is a
token, a LM assigns a probability $P(x)$.

Language models allow us to score different mutants based on the probability
they assign to different productions of code. Given the sequence of code tokens
$y_0 \ldots y_M$ the score of a mutant in location $l$ that is mutates the current
token to $t$ is given by
\begin{align}
  S(t, l) = \log\frac{P(a_0 \ldots a_M)}{P(y_0 \ldots y_M)} =
   \sum_{i=l}^{l+n} \log\left(\frac{P(a_i | a_{i-1} \ldots a_{i-n+1})}{P(y_i | y_{i-1} \ldots y_{i-n+1})}\right)
\end{align}
where $a_i=y_i$ if $i \neq l$ otherwise $a_i=t$ and $n$ refers to the size of the
\ngram LM. $S$ is the log-ratio of the probability of the mutated code to
the probability of the unmutated code and is measures how
more (or less) probable the token $t$ is in location $l$ compared to the current
token $y_l$. Since $S(t,l)$ is a log-ratio, it suggests how many orders of
magnitude more (or less) probable the mutated code is compared to the existing code.
In our work, for a given program location, we rank mutants in an ascending order
(\ie from the least natural to the most natural). Since traditional mutants do
not have a naturalness score and are a minimal set, they are placed first in
the selection list.

\subsection{Implementation}
We implemented our approaches using shared infrastructure to ensure that our results
accurately reflect differences between the mutation approaches rather than differences in
their implementations. More specifically, we developed a mutation tool that builds on top
of the Major mutation framework~\cite{just2014major}. Our tool generates traditional and
\natural mutants and collects mutation analysis data.

Our implementation that computes the control flow graph (CFG), for each method in our
dataset, relies on the abstract syntax tree (AST) representation of Eclipse's JDT. Within
such a computed CFG, each node consists of a single statement or a conditional expression
(for branches). For simplicity, our implementation treats \lstinline+try-catch+ blocks as
\lstinline+if+ statements. Some statements in a program do not belong to any explicitly
declared method but rather a (static) initializer; in such cases, our implementation
creates a separate CFG for the initializer, containing the set of sequential statements
that exist in that initializer.

\section{Evaluation}
\label{sec:eval}
This section details the evaluation of our new approaches to mutant generation and mutant
selection with respect to effectiveness and efficiency.
We determine the effectiveness of our mutant generation approach by
measuring how many mutants it generates that are \emph{coupled} with real faults,
following the approach of Just \etal~\cite{just2014mutation}.

First, we define two concepts.  When $T$ is a test suite that kills the mutant
$m$ while $T - \{t\}$ does not, then $t$ is a \emph{triggering test case} for
$m$, and $m$ is \emph{coupled} to that test. When $t$ detects the defect $d$, we
further say that $m$ is coupled to the defect $d$.

To perform this evaluation, we used Defects4J~\cite{just2014defects}.
Defects4J contains 357 defects from 5 Java programs with the state of the code
\emph{before} and \emph{after} a bug fixing commit. It additionally contains
all the unit tests of that software project. The unit test(s) that were created
or modified to detect the defect are triggering tests.
Because some mutants trigger compiler bugs, we use 258 of these
defects across all projects, spanning all projects.

\boldpara{Methodology} First we generate all mutants for all
possible program locations within the class(es) affected by the bug fixing commit. For
the traditional mutation operators, we use the operations that are commonly used
in the literature. Those are the mutation operators described in Just \etal~\cite{just2014mutation}.
A detailed list can be found in \autoref{tbl:uniqueCoupling}. These traditional
mutation operations are considered a sufficient set~\cite{offutt1996sufficient,siami2008sufficient}.
Then, for each defect, using the Major framework~\cite{just2014major} we run a full mutation
analysis executing all the non-triggering tests, collecting information about
each mutant (\ie whether it lived or was killed by the tests).
Then, we re-run a mutation analysis using only the triggering test(s).
The mutants that were killed by the triggering test(s) but not from the
non-triggering tests are the \emph{coupled} mutants for that defect.

There are various potential threats to validity. First, Defect4J contains 5 Java
programs and may not be fully representative of faults in other software
systems. However, as the creators of the dataset argue, it contains a diverse
set of software and defects. Additionally results using this data have been
correlated with other datasets~\cite{just2014mutation}.  This gives us
confidence that our results will generalize in other codebases. Another
potential threat is the use of the specific subset of traditional operators.
Although this subset can be enlarged, we follow Just \etal who argued that these
traditional mutation operators are sufficient and
effective~\cite{just2014mutation}.

\boldpara{Verifiability} Our tailored mutation operators are implemented
in the Major mutation framework~\cite{just2014major} which is
publicly released. Likewise, the Defects4J dataset is publicly available, allowing
third-party verification of the results.

\boldpara{Metrics} We are interested in the efficiency of the mutants.
We measure efficiency of a set of mutants by the number of defects
that have at least one mutant coupled with them. A mutant is coupled with
a defect if it is killed by a triggering test but \emph{not} by a non-triggering
test. The coupling suggests that the given mutant is representative and relevant to
the real defect.

We measure the effectiveness of mutants within different defect scopes. We
define the scope of a defect as the part of the code where that the bug fixing
commit made changes. In specific, we consider three increasingly narrower scopes:
\begin{squishlist}
  \item Class Scope: contains mutants only within the class(es) where the bug fixing
    commit affected.
  \item Method Scope: contains all mutants within the method(s) that
   the bug fixing commit affected.
  \item Line Scope: contains all mutants within the line(s) that
   the bug fixing commit changed.
\end{squishlist}
The notion of defect scopes is important when measuring effectiveness.
When a software engineer is interested in a specific unit (\eg
a method), she will use mutation testing only within the scope of that unit. As the
defect scope gets narrower, the mutants should be increasingly relevant to the
defect.

\subsection{Mutant Effectiveness}
\begin{figure}[t]
	\centering
	\includegraphics[width=0.99\columnwidth]{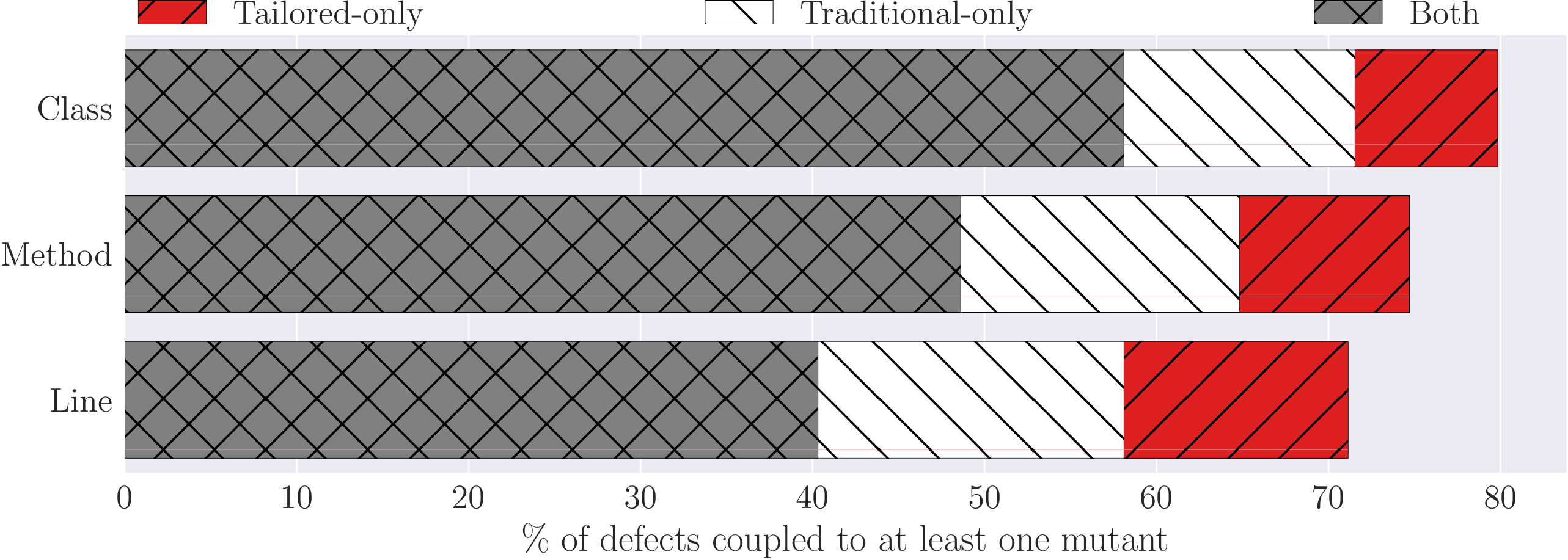}
	\caption{The percent of Defect4J defects coupled with at least one mutant at each
  scope. About 22\% of the defects are coupled with either \emph{only} tailored
  mutants or \emph{only} traditional mutants. This increases to 33\% when
  considering only mutants at the line-scope.}
	\label{fig:couplingperscope}
\end{figure}

In this section, we are interested in the effectiveness of the mutants and the
mutation operators. \autoref{fig:couplingperscope} presents the defects that
are coupled to at least one mutant at different scope. Thanks to the \natural
mutants at class scope we can detect 8\% more defects, compared to only using
traditional mutants. This increases to 14\% when considering mutants only
within the line scope of each defect. The gray area suggests that a large
number of \natural mutants are coupled with defects that are also coupled with
traditional mutants. There is also a part of traditional mutants which are
uniquely coupled with some defects. Thus, \emph{\natural mutants increase
the efficiency of mutation testing}.

Tailored mutants and traditional mutants allow us to apply mutations in different
and distinct locations. In our data for all CFG nodes that are covered by at least one mutant,
69\% of those locations are covered \emph{only} by a \natural mutant, 17\%
\emph{only} by a traditional mutant and the rest of the locations by both types of
mutants.

We are also interested in the coupling of mutation operators with real defects.
For a defect $d$, the set of coupled mutation operators is the set of mutation operators
that can produce a mutant that is coupled to $d$. Additionally, we say that
a mutation operator is \emph{uniquely} coupled if it is the only operator that
generates coupled mutants for a given defect. \autoref{tbl:uniqueCoupling}
presents the coupling statistics for the mutation operators. The
variable name replacement mutation operator achieves the highest coupling
and achieves equal unique coupling to the traditional relational operator replacement~(ROR)
operator. Tailored method replacement achieves the third best unique coupling.
\autoref{tbl:uniqueCoupling} also shows the average kill rate of the mutants produced
by each operator, using only the non-triggering tests. Averages are computed per
defect for each scope; we exclude from the average any operators that do not generate
a mutant within a defect scope.
If a mutation operator is representative for real defects, we
expect as the defect scope gets narrower, the kill rate to
decrease, implying a less tested region of code. \autoref{tbl:uniqueCoupling}
suggests that both \natural and traditional operators exhibit such behavior.
Regarding the \natural mutation operators, the larger number of generated mutants in
combination with a higher ratio of mutants detected by the non-triggering test suite
suggests that \natural mutation operators are a noisier source, which reinforces the need
for proper sampling approaches.

\begin{table*}
  \centering
  \caption{Coupled and uniquely coupled mutation operators to defects at line scope (Ordered
  by the number of uniquely coupled defects). A (uniquely) coupled
  mutation operator is an operator that generates (uniquely) coupled mutants.
  The table shows the average number of mutants (per defect)
  and the percent of mutants that were killed by the non-triggering tests at each
  defect scope. When computing the averages for an operator, we ignore defects where the operator
  could not be applied within the defect scope. } \label{tbl:uniqueCoupling}
  \begin{tabular}{llrrrrrrrr} \toprule
    &&&&\multicolumn{3}{p{2.3cm}}{Average Number of Mutants per Defect}&\multicolumn{3}{p{3cm}}{Average Kill Rate of Non-Triggering Tests}\\
    Mutation Operator & Type & Total & Unique & class & method & line & class & method & line\\ \midrule
    Variable Name Replacement (VAR) & \natural & 109 & 13 & 1512 & 184 & 48 & 77.5 & 69.0 & 52.7\\
    Relational Operator Replacement (ROR) & traditional & 76 & 13 & 127 & 18 & 6 & 62.6 & 48.7 & 40.8\\
    Method Call Replacement (MCR) & \natural & 67 & 8 & 592 & 65 & 24 & 86.3 & 72.3 & 54.5\\
    Statement Deletion (STD) & traditional& 28 & 7 & 38 & 5 & 2 & 81.2 & 71.0 & 38.5\\
    Literal Value Replacement (LVR) & traditional& 67 & 6 & 104 & 17 & 5 & 74.4 & 62.3 & 41.5\\
    Conditional Operator Replacement (COR)& traditional& 70 & 4 & 93 & 14 & 6 & 68.2 & 56.5 & 48.0\\
    Natural Literal Replacement (NLR) & \natural & 40 & 1 & 96 & 22 & 7 & 77.3 & 62.4 & 51.1\\
    Arithmetic Operator Replacement (AOR) & traditional& 25 &0 & 128 & 39& 17 & 79.1 & 65.9 & 61.5\\
    Unary Operator Replacement (ORU) & traditional& 3 & 0 & 6 & 3 & 3 & 88.6 & 77.9 & 50.0\\
    Logic Operation Replacement (LOR) & traditional& 1 & 0 & 10 & 6 & 2 & 79.6 & 92.9 & 0.0\\
    Shift Operator Replacement (SOR) & traditional& 1 & 0 & 5 & 3 & 3 & 65.8 & 75.0 & 33.3\\ \bottomrule
  \end{tabular}
\end{table*}

\boldpara{Qualitative Evaluation}
\begin{figure}[t]
	\input{figures/mutationsamples}
	\caption{Snippets of real faults and their fixes in Defect4J coupled \emph{only} to
  \natural mutants. We also present sample of the tailored mutations that
  detect the fault. The formatting of some snippets has been changed to fit the page.}
	\label{fig:qualitative}
\end{figure}
Tailored mutants mutate the code in ways that are in some cases more \natural
and related the original defect. \autoref{fig:qualitative} presents a sample
of the \natural mutants coupled at the line scope where no traditional
mutant was coupled within that scope. Variable name replacements (\textsf{Math 70},
\textsf{Time 21} in \autoref{fig:qualitative}) capture confusions of variable
usage, that sometimes have similar names and function. Method call replacement
is able to introduce bugs that may arise from faulty usage of similar functions.
All such mutations were not previously possible with the traditional set of
mutation operators.

\begin{figure}[t]
	\input{figures/failedexamples}
	\caption{Snippets of real faults and their fixes (in Defect4J) with no
  coupled mutants that could be detected
  by other kinds of tailored mutation operators. The formatting
  of some snippets has been changed to fit the page.}
	\label{fig:failed}
\end{figure}
\autoref{fig:failed} shows some examples of real faults that are not detected
by either natural or traditional mutants, but could be detected by other unexplored
kinds of tailored operators. Mutation operators that refine conditions with
special cases (\textsf{Closure 102}), argument shuffling (\textsf{Time 4}),
change of types when implicit type conversion may be happening (\textsf{Math 30}),
statement location perturbation (\textsf{Closure 102}) and other complex
pattern matching conditionals (\textsf{Closure 112}) are some of the classes
where tailored mutation operators are needed to avoid an exponential explosion
of mutants. Future research is required to investigate if code naturalness or
other methods can provide a reasonable set of such operators.

\boldpara{The effectiveness of natural literal mutation} Natural literal
mutation approaches the problem of mutating variables and literals into
other literals in a significantly different way compared to traditional
literal value replacement operators. There is an unbounded
number of literals to which a primitive variable or literal can be mutated.
Natural literal replacement operators take a project-specific, context-specific approach to expand
the traditional literal mutations by
suggesting mutations that make the code similar to other code locations
that have appeared within the project. The effectiveness of the natural literal
replacement operation is indicated by the fact that 15.8\% of defects
are coupled to at least one natural literal mutation. This number increases
to 22.9\% and 28.9\% when considering mutants within method and
line scope.

For example, in \textsf{Lang 41} the string literal ``\lstinline+[]+'' is mutated to
``\lstinline+[+'' within the code \lstinline+arrayPrefix.append("[]");+ because it has
been seen within similar context. This mutation produces a mutant that is coupled
with a real defect at the line scope. However, natural literal mutations have fewer 
uinquely coupled defects than
the traditional literal mutation (\autoref{tbl:uniqueCoupling}). We believe that
this is because some contexts are relatively unique and infrequent within a single
project and no alternative literals can be suggested. Investigating expanding
literal mutations from a larger corpus of code is left for future work.
Although literal mutations do not seem to be as effective on their own as traditional literal
value replacements, they still seem to be a good complement for increasing the breadth
of literal mutations and their overall efficiency.

\subsection{Mutant Selection}
The increased number of mutants raises the problem of selecting mutation
operations while maintaining the efficiency of the mutation analysis. This
section evaluates various approaches for mutant selection. As a baseline
we use a random selection policy, where we pick a mutant at random. In this
case, the expected effectiveness can be computed analytically by calculating the probability
that at least one coupled mutant has been killed by the triggering test(s), which
is given by
\begin{align}
  P(\kappa, \lambda, |\mathbb{M}|) =
  \begin{cases}
    1       & \quad \text{if } |\mathbb{M}| - \kappa < \lambda \\
    1-\frac{(|\mathbb{M}|-\kappa)!(|\mathbb{M}|-\lambda)!}{|\mathbb{M}|!(|\mathbb{M}|-\kappa-\lambda)!}  & \quad \text{otherwise}\\
  \end{cases}
\end{align}
where $\mathbb{M}$ is the set of all available mutants in the scope, $\kappa \leq |\mathbb{M}|$
is the number of mutants we are currently selecting and
$\lambda$ is the number of mutants in $\mathbb{M}$ that are killed only by
the triggering test (\ie are coupled mutants).

\begin{figure*}[t]
	\centering
  \begin{subfigure}[b]{0.495\textwidth}
        \includegraphics[width=\textwidth]{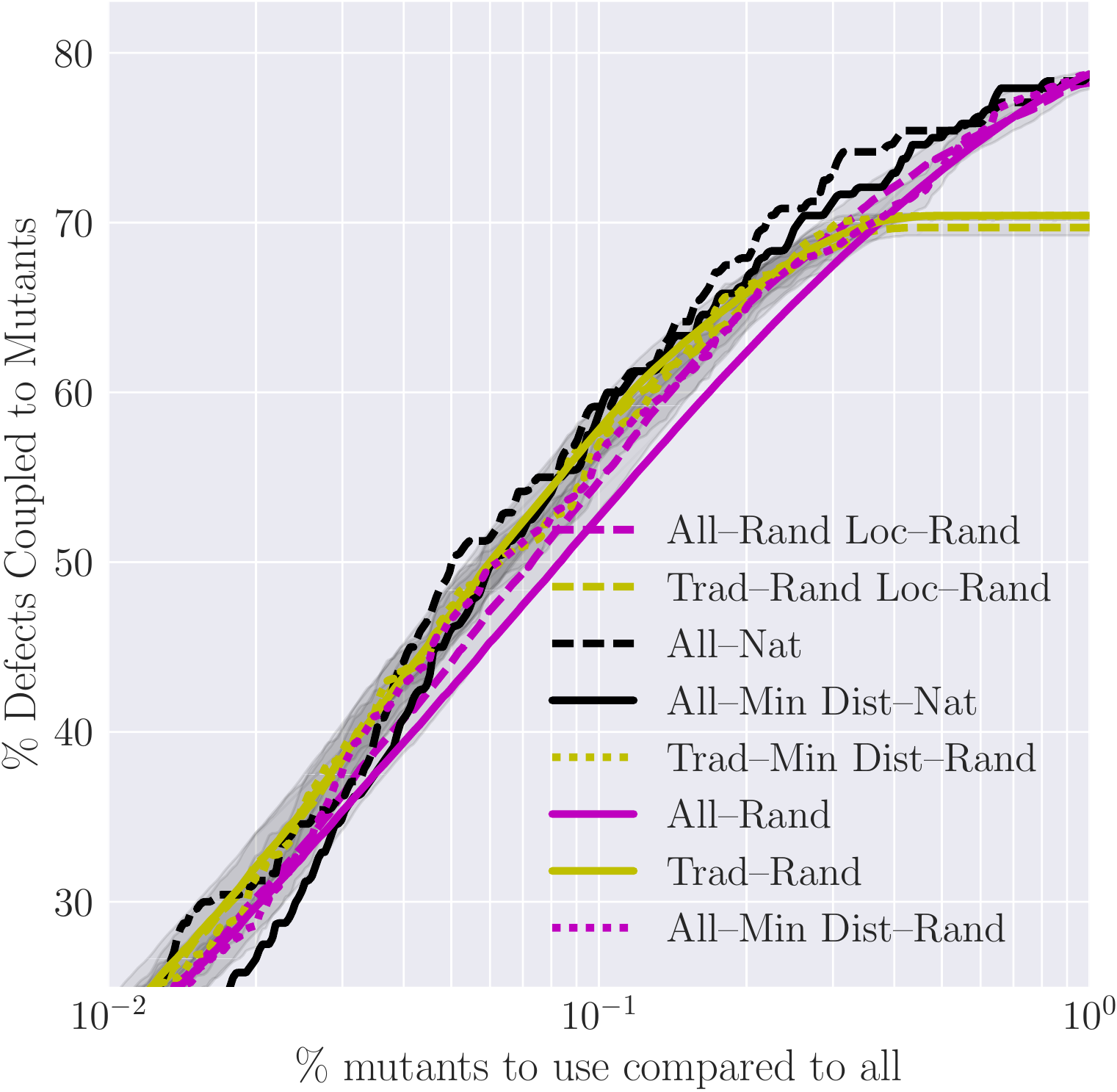}
        \caption{Class Scope}\label{fig:effectivenessGeneralClass}
  \end{subfigure}
  \begin{subfigure}[b]{0.495\textwidth}
        \includegraphics[width=\textwidth]{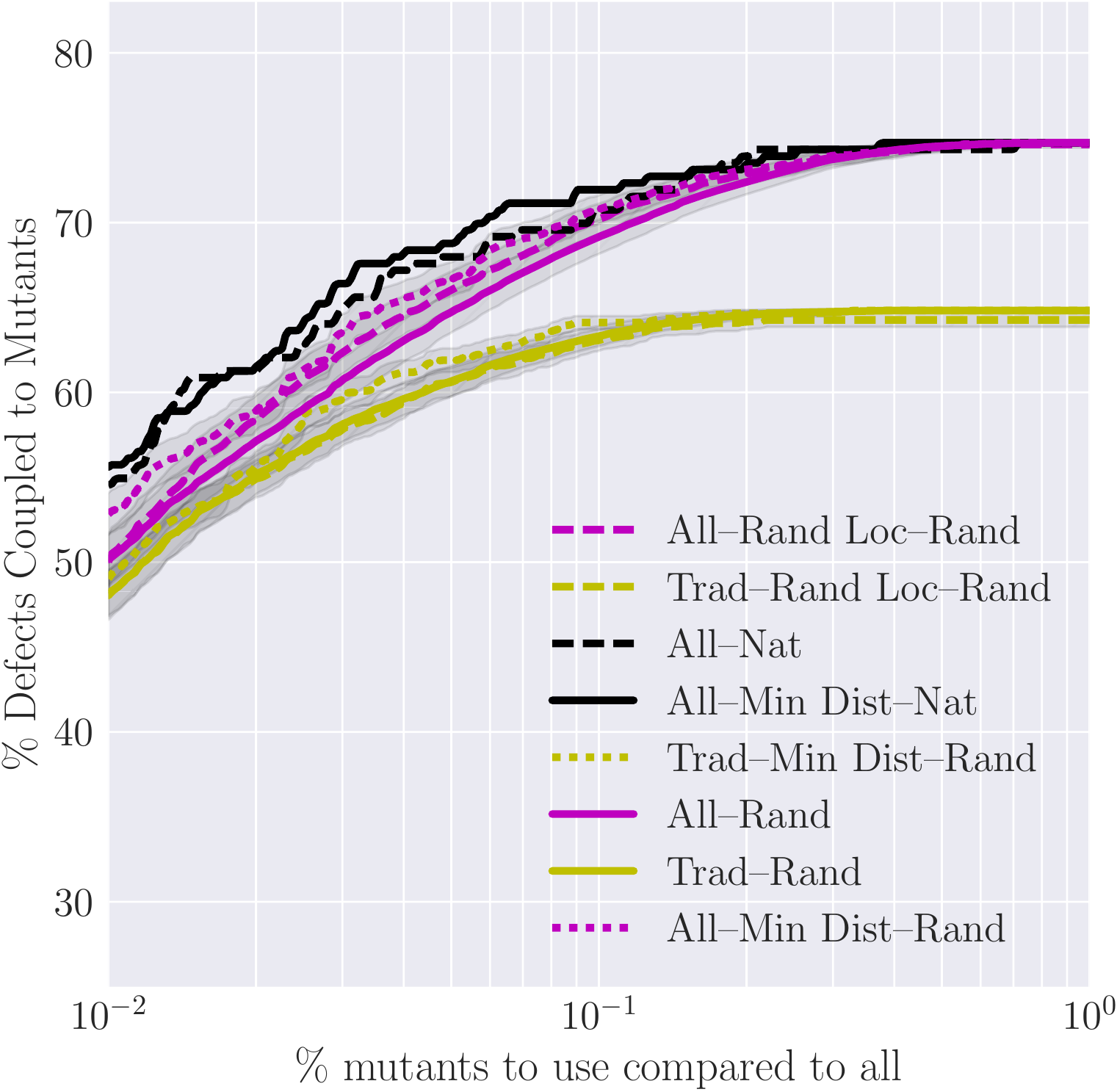}
        \caption{Method Scope}\label{fig:effectivenessGeneralMethod}
  \end{subfigure}
	\caption{Mutant effectiveness of traditional (``Trad'') and all (``All'') mutants.
  The $x$ axis (in log-scale) is the mutant budget (in relative
  terms to the total number of \natural and traditional mutants of each defect)
  that is executed during mutation analysis. One
  may notice that when we use both traditional and tailored mutants at the class scope,
  the effectiveness is increased but with the extra cost of executing
  more mutants. However, when limiting ourselves to the method scope of a defect, including tailored mutants
  always helps. The graphs show different methods for selecting
  mutants. For the selection processes that contain stochasticity (``Rand''), the
  shaded region around each line indicate one standard deviation from
  the expected efficiency. Selection processes that use our submodular objective
  are denoted by ``Min Dist''. We do \emph{not} show results for line defect scope since
  selection methods behave in a very similar way because there are few mutants
  within a given line of code. }
	\label{fig:mutantEffectivness}
\end{figure*}

\autoref{fig:mutantEffectivness} plots the effectiveness of the mutants as
more mutants are selected with different selection policies. As previously discussed,
selecting both \natural and traditional mutants achieves the highest efficiency ($y$-axis).
However, this requires to incur the cost of potentially executing more mutants. \autoref{fig:effectivenessGeneralClass}
suggests that for a small numbers of mutants --- within class scope --- just
using traditional mutants is the most efficient possible choice. However, as
the budget of mutants to execute becomes larger, using both \natural and traditional mutants
achieves the best effectiveness.
This result holds \emph{only} when
considering class-level defect scope. When we narrow the scope (\ie consider mutants
only within the method or line scope) a na\"ive random selection of both \natural and traditional mutants
seems (\autoref{fig:effectivenessGeneralMethod} for method scope, but the results
are similar within line scope) to always provide better effectiveness, regardless
of the mutant selection method.
This results suggest an actionable recommendation for practitioners using
mutation analysis. \emph{When interested in a single method (or line) using both
tailored and traditional mutants provides better efficiency irrespectively of
the number of mutants that one may wish to execute.}

\begin{figure}[t]
	\centering
	\includegraphics[width=0.99\columnwidth]{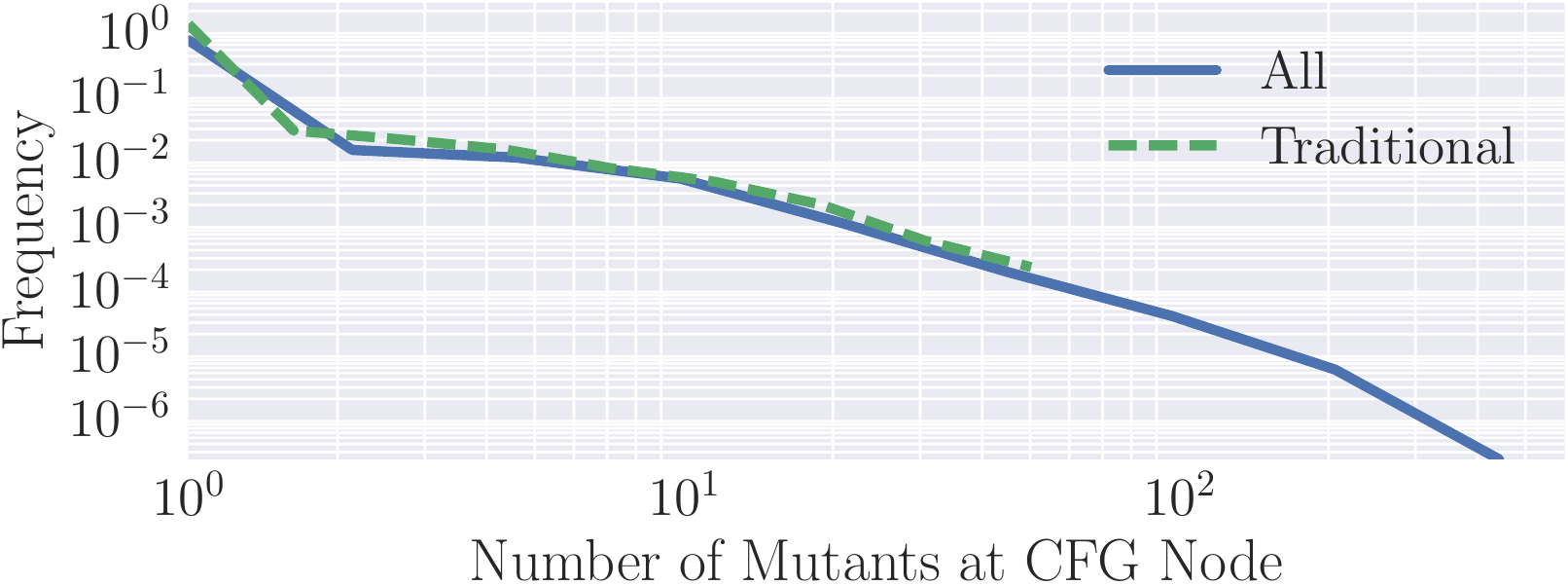}
	\caption{The distribution of number of mutants per control
  flow graph (CFG) node. Each node in the CFG is either an executable or non-executable statement.
  About 85\% of the CFG nodes are associated with at least
  one mutant. The graph shows a heavy-tailed distribution
  of mutants to locations, showing that few nodes
  have hundreds of mutants, while most of the nodes have only
  a few. Note the log-log scale, which is customarily used to
  plot such distributions.}
	\label{fig:operatorLocationDistribution}
\end{figure}

\boldpara{Importance of Location in Mutation Testing}
Mutation location plays an important role in mutation testing, since it is
correlated with various coverage metrics. In the Defects4J dataset,
we find that the mutants are \emph{not} evenly distributed across locations
in the control flow graph (CFG) of each program. \autoref{fig:operatorLocationDistribution}
shows the distribution of the number of mutants per CFG node. We find that
some CFG nodes have hundreds of mutants, while others have only a few. We
further note that although we have introduced the \natural
mutants, still about 15\% of the CFG nodes are \emph{not} associated with
any mutant.

\begin{figure*}[t]
	\centering
  \begin{subfigure}[b]{0.325\textwidth}
        \includegraphics[width=\textwidth]{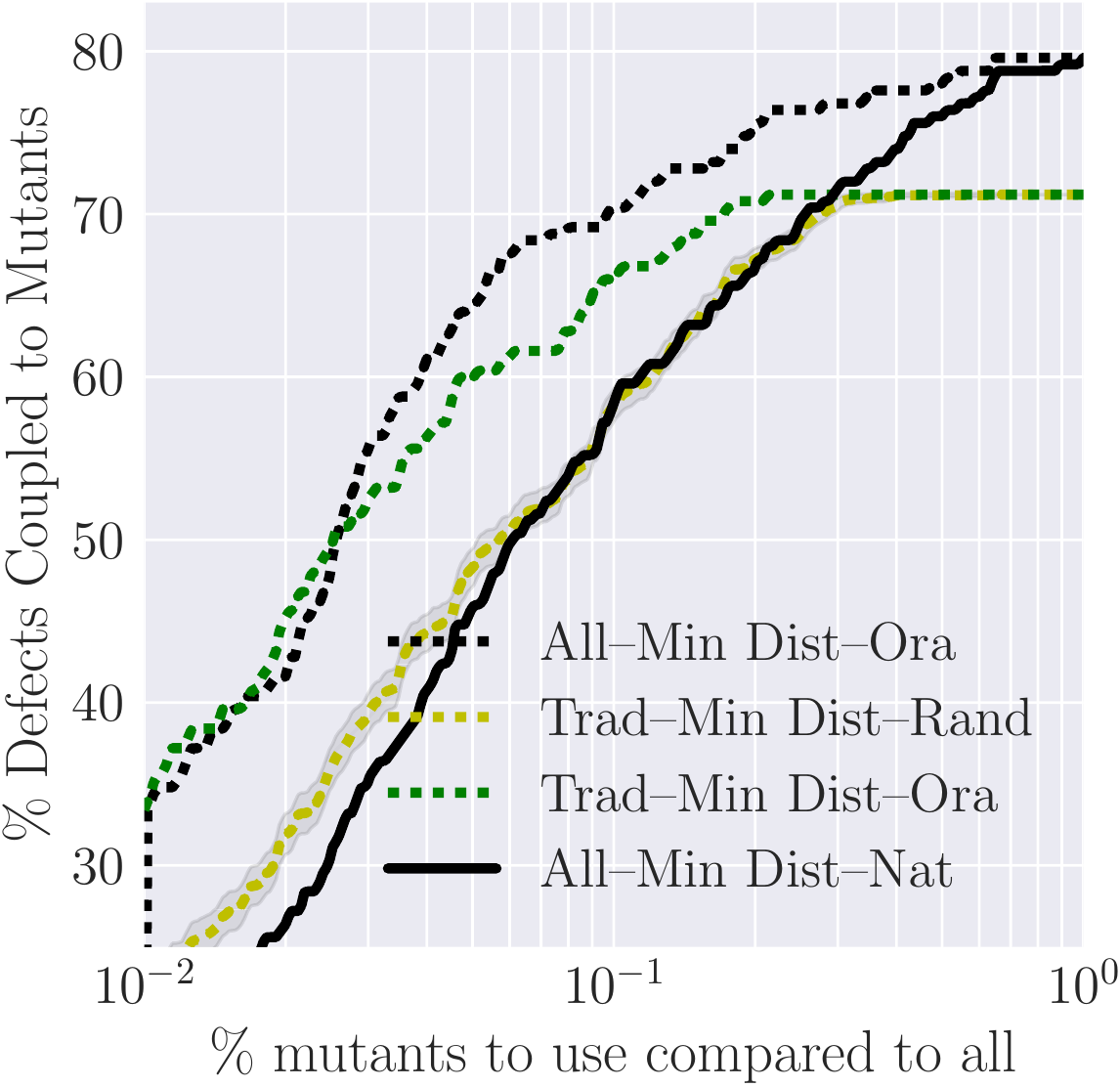}
        \caption{Class Scope}\label{fig:oracleClass}
  \end{subfigure}
  \begin{subfigure}[b]{0.325\textwidth}
        \includegraphics[width=\textwidth]{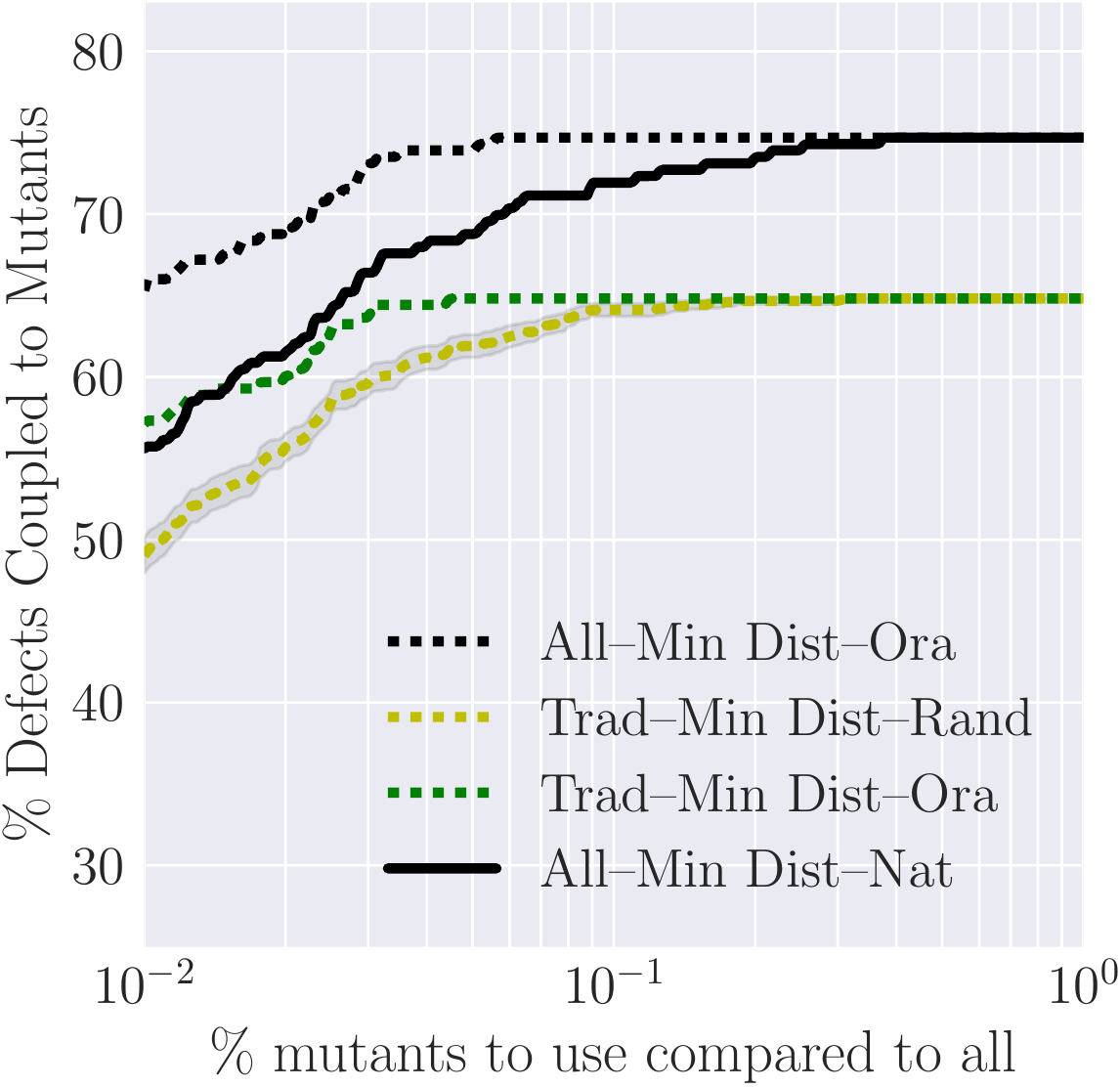}
        \caption{Method Scope}\label{fig:oracleMethod}
  \end{subfigure}
  \begin{subfigure}[b]{0.325\textwidth}
        \includegraphics[width=\textwidth]{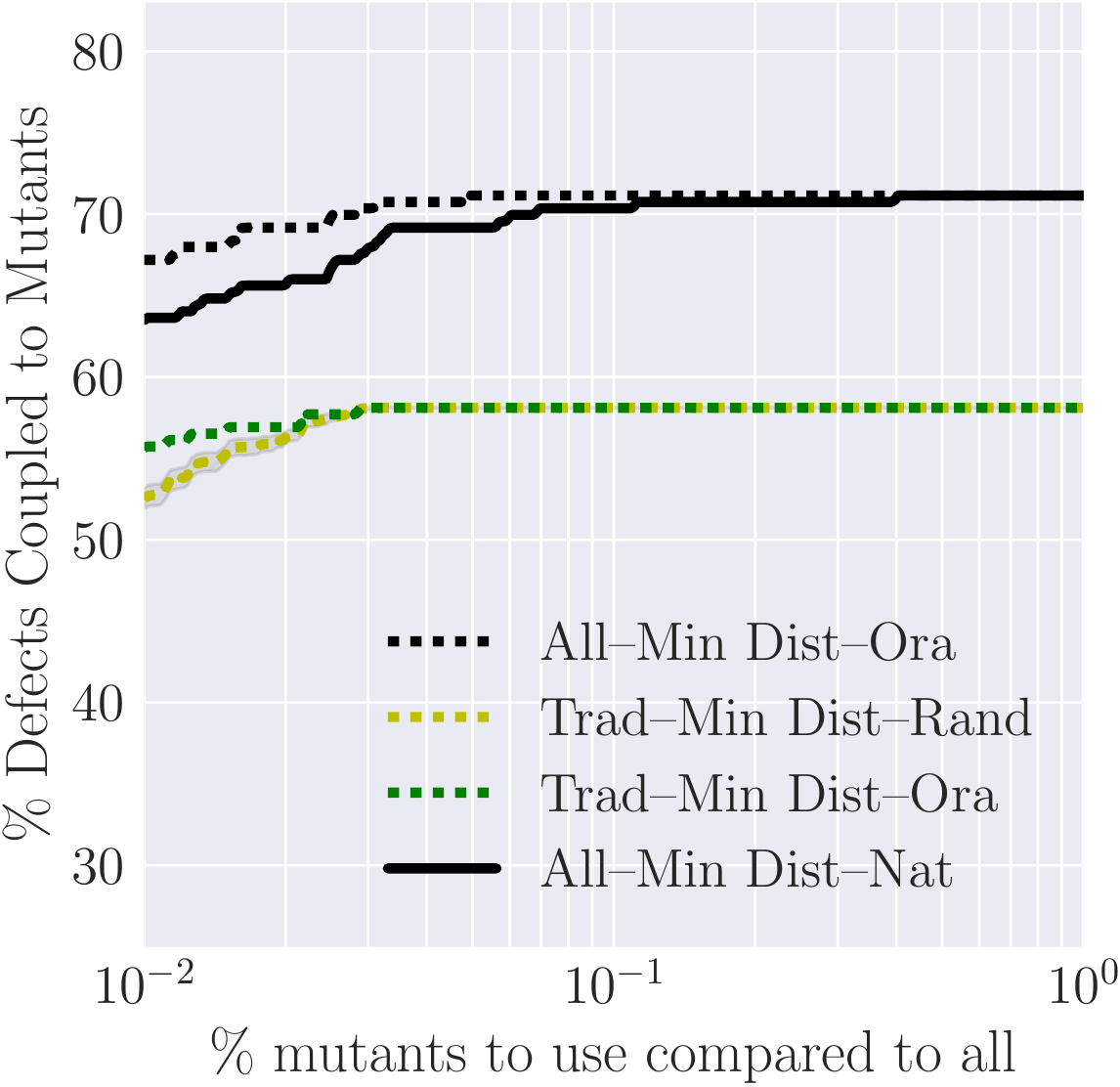}
        \caption{Line Scope}\label{fig:oracleLine}
  \end{subfigure}
  \caption{Mutant effectiveness of traditional (``Trad'') and all (``All'') mutants for a given budget.
  The $x$ axis (in log-scale) is the budget (in relative
  terms compared to using all \natural and traditional mutants of a defect)
  that are executed during mutation analysis, as in \autoref{fig:operatorLocationDistribution}.
  The graphs show the performance of the location-based selection for traditional and
  all mutants when selecting mutants with out optimization-based approach that
  minimized \autoref{eq:distanceOptimization}. For each defect scope, we show
  the performance of the method when using code unnaturalness \vs using an oracle
  to select the optimal mutant at each location. The oracle is the upper bound that the CFG distance
  minimization can achieve.}
	\label{fig:oracleSelection}
\end{figure*}

\boldpara{Comparing Effectiveness of Selection Policies}
In \autoref{fig:mutantEffectivness}, we plot the percent
of mutants that one can use ($x$-axis) compared
to using all the available mutants \vs the
percent of bugs that are coupled with at least one of the selected mutants.
The $x$-axis (in log scale), essentially
can be interpreted as the budget of mutant executions that one may be willing
to incur during mutation analysis.

\autoref{fig:mutantEffectivness} show two trends. First, location-based methods
perform better than a simple baseline random selection. Additionally, there is
an improvement when using our CFG distance optimization and
our naturalness-based scoring, compared to a random selection for a given location.
The differences seem to diminish when considering only line scope (not shown). This
is reasonable since the number of locations within a line are quite limited
(most usually we would have exactly one location). We also observe that our
selection method achieves better results within a method scope. This implies that within
the class scope we have a suboptimal selection method (\autoref{eq:distanceOptimization}
implies that methods with more CFG nodes are prioritized) and that further research
is required for selecting the methods first.

The goal of selecting mutants is to achieve a similar efficiency with a
smaller number of mutants. Our selection methods achieve this. For example,
in \autoref{fig:effectivenessGeneralMethod} to achieve 70\% of efficiency, one
needs to use both \natural and traditional mutants. If a developer resorts to simple
random selection, then she will need to execute 13\% of all mutants. In contrast,
only 6\% of all mutants are needed when preferring the
mutants returned by our selection method that combines our location optimization
and naturalness-based selection, which reduces the number of mutants that need
to be executed by about 50\%. Similarly, for an effectiveness of 60\% our
selection method requires about half the mutants compared to a na\"ive random
approach. These results suggest that location and naturalness-based selection
of mutants is a promising approach for reducing the number of mutants required
to achieve a given effectiveness.

\boldpara{The upper bound of location-based selection}
\autoref{fig:oracleSelection} plots the effectiveness of the best selection
processes previously discussed. The graphs
also include an ``oracle'' selection process. This refers to
a selection process, where we first pick a location optimizing the objective
in \autoref{eq:distanceOptimization} and then at each location an oracle
picks the best possible mutant (\ie if a coupled mutant exists, it is picked.
Otherwise a random non-coupled mutant is selected). Obviously, this is \emph{not}
a real-life option but it provides an upper bound on the efficiency of our
location-based optimization selection method. The gap seen in \autoref{fig:oracleSelection} suggests
that improvements in the mutant selection at a given location can yield
substantial improvements to the efficiency of the location-based methods. It also suggests
that source code naturalness is only one indicator of the effectiveness
of a mutant. We believe that this has to do with two factors. First, simple
errors that may be captured by a local model (such as the \ngram) are most
often easily removed before the code is even committed or tested. The remaining
bugs, should contain more semantics-related which is harder to capture with an
\ngram LM. Second,
it could be that a large number of defects that are too
unnatural represent bugs that are caught early in the development process.

\section{Related Work}
\label{sec:relwork}

\boldpara{Mutation Analysis}
The huge computational cost of mutation analysis is a widely acknowledged and studied
problem. Selective mutation testing, first studied by Mathur~\cite{mathur1991selective},
aims to select a subset of all mutants that is representative of the entire set. A large
body of related work primarily focused on the determination of sufficient mutation
operators in an attempt to reduce the number of generated mutants. Wong and
Mathur~\cite{wong1995selective} found that mutants generated with only 2 out of 22
mutation operators achieved effectiveness results that are comparable to the full set of
mutants. Offutt \etal~\cite{offutt1996sufficient} determined a sufficient set of 5
mutation operators that only incurred a negligible loss in effectiveness in their
experiments. Similarly, Namin \etal~\cite{siami2008sufficient} analyzed an exhaustive set
of 108 mutation operators and concluded that 28 are sufficient. More recently, however,
Zhang \etal~\cite{zhang2010selective} showed that random selection of mutants is as effective
as operator-based mutant selection, and a study by Kurtz \etal~\cite{kurtz2016selective}
suggests that there exists no subset of mutation operators that can generate effective
mutants for all programs.

Our work is motivated by recent findings and differs from the large body of related work
in two ways. First, rather than sampling from a vast, highly redundant pool of mutants,
our approach starts with a small set of provably effective and
non-redundant~\cite{kaminski2013redundant,just2015redundant} mutation operators, and
augments it with mutation operators tailored towards the program under test. Second, prior
work has evaluated the effectiveness of a sampling approach exclusively on mutants whereas
our methodology considers the coupling of mutants with real faults.

Prior work has intensively addressed the scalability problems of mutation analysis. In
contrast, the effectiveness, \ie, the correlation between mutant detection and real fault
detection, is less studied. Andrews \etal~\cite{andrews2005mutation} and Just
\etal~\cite{just2014mutation} showed that such a correlation indeed exists but the latter
study also showed that the set of commonly used mutation operators is not
sufficient---27\% of real faults are not coupled to the mutants they generate. Our work
aims to close this gap by providing tailored mutants that are coupled to more real faults.

\boldpara{Code Naturalness}
Recently, machine learning and NLP methods have been applied to source code
text with fruitful applications. Originally,  Hindle \etal~\cite{hindle2012naturalness} showed
that source code is predictable and ``natural'', by using \ngram LMs. Later,
others~\cite{allamanis2013mining,maddison2014structured,nguyen2013statistical,tu2014localness}
made improvements to the language models of code. Applications of these models
include source code autocompletion, learning coding conventions and
suggesting names~\cite{allamanis2014learning,allamanis2015suggesting},
predicting program properties~\cite{raychev2015predicting},
extracting code idioms~\cite{allamanis2014idioms},
code migration~\cite{karaivanov2014phrase,nguyen2014migrating}
and code search~\cite{allamanis2015bimodal}.
Related to our work is the work of Cambell \etal~\cite{campbell2014syntax}
and Bhatia and Singh~\cite{bhatia2016automated}
that provide better localization of compile-time syntax-only errors using code language
models. Recently, Ray \etal~\cite{ray2015naturalness} showed that
buggy code has on average lower entropy scores assigned by cache \ngram models.

\section{Conclusion}
\label{sec:conc}

In this work, we introduce the idea of tailored mutation operators, which
extend traditionally used mutation
operators by adapting the code transformations they produce
using information from elsewhere in the project. We
presented  examples of tailored mutation operators that replace variable names, method calls
and literals. These new operators increase the coupling of mutants to real defects compared to traditional operators.
Introducing \natural mutants increases effectiveness but increases the number
of mutants that need to be executed. To alleviate this problem, we present
 new mutant selection methods. We select program locations within
a control flow graph first by employing submodular optimization, a well studied theoretical framework for optimization over sets,
and then we select mutants within that location, based
on the naturalness of the mutation, following recent work
that shows that buggy code is likely to be \emph{un}-natural
\cite{ray2015naturalness}.

Promising future directions include the exploration of alternative methods
for selecting mutants within a given program location. Additionally, using
more advanced models of code naturalness may improve upon the performance
of the \ngram language model. Modification to the location selection method
and its submodular optimization approach, \eg using machine learning, may provide additional improvements.

\section*{Acknowledgments}
  This work was supported by Microsoft Research through
  its PhD Scholarship Programme.  Charles Sutton and Miltiadis Allamanis
  were supported by
  the Engineering and Physical Sciences
  Research Council [grant number EP/K024043/1].

\bibliographystyle{IEEEtran}
\bibliography{lit/bibliography,lit/mutation.bib} \balance
\end{document}